\documentclass[pre,floats,aps,amssymb,preprint,epsf,rotate]{revtex4}

\usepackage[dvips]{graphicx}

\begin{document}

\title{Ballistic deposition on deterministic fractals:
On the observation of discrete scale invariance} 

\author{Claudio M. Horowitz $^{1}$, 
Federico Rom\'a $^{2}$ and Ezequiel V. Albano $^{1}$.}

%\address
\affiliation{ 1- Instituto de Investigaciones Fisicoqu\'imicas Te\'oricas 
y Aplicadas, (INIFTA), UNLP, CCT La Plata-CONICET, Sucursal 4, Casilla de Correo 16, (1900)
La Plata. Argentina \\
2- Centro At\'omico Bariloche, 8400 S. C. de
Bariloche, R\'io Negro, Argentina}
%\address{1- Departamento de F\'{\i}sica, Universidad Nacional
% de San Luis, Chacabuco 917, CP 5700, San Luis, Argentina. }  

\date{\today}

\begin{abstract}

The growth of ballistic aggregates on deterministic fractal substrates is 
studied by means of numerical simulations. First, we attempt the 
description of the evolving interface of the aggregates by applying the 
well-established Family-Vicsek dynamic scaling approach. Systematic 
deviations from that standard scaling law are observed, suggesting that
significant scaling corrections have to be introduced in order to achieve 
a more accurate understanding of the behavior of the interface. Subsequently,
we study the internal structure of the growing aggregates that can be 
rationalized in terms of the scaling behavior of frozen trees, i.e., 
structures inhibited for further growth, lying below the growing interface.
It is shown that the rms height ($h_{s}$) and width ($w_{s}$) of the trees of
size $s$ obey power laws of the form $h_{s} \propto s^{\nu_{\parallel}}$ and  
$w_{s} \propto s^{\nu_{\perp}}$, respectively. Also, the tree-size distribution 
($n_{s}$) behaves according to $n_{s}\sim s^{-\tau }$. Here, $\nu_{\parallel}$ and 
$\nu_{\perp}$ are the correlation length exponents in the directions parallel 
and perpendicular to the interface, respectively. Also, $\tau$ is a critical 
exponent. However, due to the interplay between the discrete scale invariance 
of the underlying fractal substrates and the dynamics of the growing process, 
all these power laws are modulated by logarithmic periodic oscillations.
The fundamental scaling ratios, characteristic of these oscillations, can be 
linked to the (spatial) fundamental scaling ratio of the underlying fractal
by means of relationships involving critical exponents. We argue that the 
interplay between the spatial discrete scale invariance of the fractal
substrate and the dynamics of the physical process occurring in those media 
is a quite general phenomenon that leads to the observation of 
logarithmic-periodic modulations of physical observables.           

PACS numbers: 68.35.Ct, 05.45.Df, 02.50.-r, 81.15.Aa

\end{abstract}

\maketitle

\newpage

\section{Introduction}

The study, characterization, and understanding of growth processes that take
place under far-from equilibrium conditions are topics that have attracted 
great attention due to their relevance in many fields of science and 
technology \cite{BS,2,3,4}.
Growing aggregates of biological origin (bacteria, fungi, tumors, etc.),
the deposition of thin films, the growth of magnetic materials, alloys and 
polymers, among others, have recently become the subject of 
extensive studies \cite{BS,61,6,8,5}. 
Also, the dynamic evolution of interfaces is closely related to almost all 
growth processes \cite{BS}. The characterization of the properties of interfaces has 
achieved considerable progress during the last two decades, mostly due
to the success of the concepts of the dynamic scaling theory developed by 
Family and Vicsek \cite{5b,5a}, which allows a comprehensive description of 
interfaces in terms of universality classes that group systems described 
by the same set of physically meaningful exponents.        

In order to provide a more complete description of growing aggregates, it is
desirable not only to focus on the properties of the evolving interface,
but also to attempt the simultaneous characterization of bulk properties.
In fact, in some cases, growing processes lead to the formation of porous 
materials that inherently have very interesting physical and chemical properties
with many potential practical applications.

Within this broad context, ballistic deposition (BD), which was originally 
proposed by Vold \cite{BD} as a model for the description of sedimentary 
rock formation, has become an archetypical system for the study of growing 
aggregates \cite{BS}. BD aggregates are characterized by a porous structure
in the bulk and a rough evolving interface. The interface roughness of
BD has extensively been studied by means of computer simulations and analytical
approaches in connection to the Kardar-Parisi-Zhang (KPZ) theory \cite{BS,KPZ}.

In spite of the considerable effort devoted to understanding growing
aggregates in Euclidean substrates \cite{BS,2,3,4}, to the best of our
knowledge little attention has been drawn to the study of deposition models on fractal media.
It should be expected that the interplay between the self-similarity of the
substrates and the growing mechanisms would lead to the formation of
interesting and complex porous (bulk) structures. Also, the fractality of the substrate
would affect the self-affine nature of the growing interface. So, it would
not be surprising to find nonvanishing scaling corrections to the
well-established phenomenological dynamic approach of Family-Vicsek \cite{5b,5a,BS}.
In fact, in some pioneering works \cite{bessis,brigita54} and in more recently 
ones \cite{we2,we,lett} it has shown that the power-law behavior of some observables becomes modulated by logarithmic oscillations due to the effect of the underlying lattice structure.
Also, in the last years two interesting books have been published where this issue is addressed \cite{libro1,libro2}.

Within this context, the aim of this paper is to study the scaling behavior of 
the BD model on deterministic fractal substrates. The study is based on Monte Carlo 
numerical simulations analyzed by means of the Family-Vicsek phenomenological 
scaling approach \cite{5b,5a} and on extensive numerical investigations of 
the scaling behavior of the internal structures of the growing 
system \cite{RV,Meakin0,zinc,Meakin,trees,arpascu}.
For this purpose the manuscript is organized as follows: firstly, in Section II
we provide a brief theoretical background on the scaling behavior of growing aggregates. 
The fractal substrates used for the growth of BD aggregates and
the concepts of space and time discrete scale invariance are described and 
discussed in Section III.
Subsequently, in Section IV, we describe the BD model on fractal media, as well as  
the determination of the internal structure of the aggregates in terms of frozen 
and growing trees. The results obtained by applying the standard dynamic scaling 
approach to the data are presented and discussed in Section V, while 
the analysis of the data of the internal structure of the aggregates 
is performed in Section VI.
Finally, our conclusions are stated in Section VII.

\section{Brief theoretical background}

Models aimed to describe growing aggregates may be defined and studied by means
of both continuous approaches, which involve the formulation of analytical 
equations, and discrete lattice models, which consider the deposition 
of individual  particles.
A discrete model is defined by means of a set of deposition rules that provides a 
detailed microscopic description of the evolution of the aggregate. In
these discrete models, the growing interface of the aggregate is described 
by a discrete set ${h(i,t)}$, which represents the height of site $i$ at time $t$. 
The interface then has $L^d$ sites, where $L$ is the linear size and $d$ is 
the dimensionality of the substrate (as usual, $d$ is assumed to be an integer). 
 
The dynamic evolution of the aggregate interface is characterized through 
the scaling behavior of the interface width $W(L,t)$, given by

\begin{equation} 
W(L,t)\equiv 
\sqrt{1/L^{d}\sum_{i=1}^{L^d}[h(i,t)-\left\langle  h(t)\right\rangle ]^2} .
\label{anchoto}
\end{equation} 

\noindent For this purpose, the Family-Vicsek phenomenological scaling approach 
\cite{5b,5a}, which has proved to be very successful, can be written as 

\begin{equation} 
W(L,t) =  L^{\alpha} W^{*}(t/L^{z}) ,
\label{FV}
\end{equation} 

\noindent where $W^{*}$ is a scaling function. In fact, it may be expected 
that $W(L,t)$ will show the spatiotemporal scaling behavior given by \cite{5b,5a}: 
$W \propto L^{\alpha}$ for $t \gg t_{c}$ and $W(t) \propto t^{\beta}$ for 
$t \ll t_{c}$, where $t_{c} \propto L^{z}$ is the
crossover time between these two regimes. The scaling exponents $\alpha$,
$\beta$, and $z = \alpha/\beta$ are called roughness, 
growth and dynamic exponents, respectively. Also, different
models can be grouped into universality classes when they share the 
same scaling exponents. 

An alternative method that can also be used for the characterization of 
interfaces is related to the description of the internal structure of the growing
system. This approach is based on the fact 
that any growing system can effectively be rationalized on the basis
of a treeing process, i.e., any growing structure can be thought as 
the superposition of individual trees \cite{RV,Meakin0,zinc,Meakin,MM}.
Those trees that spread out incorporating additional growing centers, e.g., 
capturing particles, developing new branches, are said to be alive. In contrast, 
other trees that may stop growing due to shadowing by surrounding growing trees 
are termed dead trees. The structure of dead trees remains frozen because it 
cannot be modified by any further growth.  
It is well known that for growing aggregates on substrates having integer 
dimension, the rms height ( $h_{s}$ ) and the rms width ( $w_{s}$ ) 
of dead trees of size  $s$ ($s$ is the number of particles belonging to the tree)
obey simple power laws given by  

\begin{eqnarray}
h_{s} \propto s^{\nu_{\parallel}}  
\label{Eq.1}  
\end{eqnarray}  

\noindent and
\begin{eqnarray}
w_{s} \propto s^{\nu_{\perp}},  
\label{Eq.2}
\end{eqnarray}  

\noindent where $\nu_{\parallel}$ and $\nu_{\perp}$ are the
correlation length exponents parallel and perpendicular to the main growing
direction of the aggregate \cite{RV,Meakin0,zinc,Meakin,arpascu,MM}, respectively.

By assuming both the proportionality between the  correlation length perpendicular
to the main growing direction and $w_{s}$, as well as $h_{s} \propto t$, one has that 
the dynamic exponent $z = \alpha/\beta$ is given by \cite{Krug} 

\begin{eqnarray}
z = \nu_{\parallel}/\nu_{\perp}. 
\label{Eq.8}
\end{eqnarray}

Furthermore, one also expects that during the competition among trees along the
evolution of the aggregate, the existence of large neighboring trees may inhibit 
the growing of smaller ones. This competing process ultimately leads to the death of some
trees that become frozen within the underlying aggregate. These prevailing
large trees continue the competition within more distant trees in a dynamic
process. Since this situation takes place on all scales, it is reasonable to
expect that the tree size distribution ($n_s$) should also exhibit a 
power-law behavior, so that \cite{RV,Meakin0,zinc,Meakin,MM,Krug}

\begin{eqnarray}
n_{s}\sim s^{-\tau },  
\label{Eq.3}
\end{eqnarray}    

\noindent where  $\tau$ is an exponent.

\section{The fractal substrates and discrete scale invariance}

In the present paper we used Sierpinski carpets (SC) as substrates for the 
growth of ballistic aggregates. In fact, SC's provide generic models for the 
building of both deterministic and nondeterministic fractals. In order to 
generate a SC embedded in $d = 2$ dimensions, one proceeds  as follows: 
a square is divided into $l^{2}$ subsquares, and then $(l^{2}-N_{occ})$ 
subsquares are deleted from the initial square ($N_{occ}$ is the number of occupied 
subsquares). This process is iterated in the remaining subsquares $k$ times, 
where $k$ accounts for the number of different generations. If the deleted 
subsquares are chosen in the same way in all iterations, the resulting fractal 
is deterministic, but if the deleted subsquares are selected at random, one 
generates a nondeterministic fractal. The mathematical fractal, obtained in 
the limit $k \rightarrow \infty $, is generically called $SC_{x}(l,N_{occ})$.
Also, the fractal associated with a finite number of segmentation 
steps is denoted by $SC_{x}(l,N_{occ},k)$. In both cases the index $x$ refers 
to the topological features of the generating cell ($SC_{x}(l,N_{occ},1)$).
The size of the lattice, where the finite fractal is embedded, then is $L = l^{k}$.

\begin{figure}[h]
\includegraphics[width=15.0cm,clip=true,angle=0]{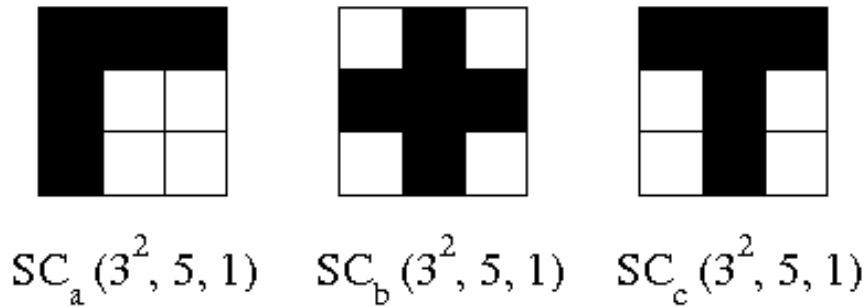}
\caption{Sketches of the three different generating cells
used to grow ballistic aggregates by keeping $l = 3$ and $N_{occ} =5$.}
\label{carpetas}
\end{figure}

For deposition models grown in integer-dimensional substrates one always has 
that all fragments or parts of the aggregate are connected to each other
through paths of nearest-neighbor occupied sites. So, it is no
longer possible to have isolated fragments of the aggregate on the substrate.
On the other hand, for some fractal substrates with noninteger fractal
dimension, it could be possible to observe the formation of isolated fragments
of the aggregate, and consequently in order to avoid this shortcoming, 
one has to carefully select suitable substrates. So, this point is essential 
for the choice of the fractals that can be used in order to study ballistic 
deposition on this kind of substrate. Accordingly, 
in this paper we use deterministic fractal substrates generated by taking 
$l=3$ and $N_{occ}=5$, and in order to prevent the fragmentation 
of the aggregates, as mentioned above, and to account for the usual 
requirement of periodic boundary conditions, 
one ends up with only three generating cells, as shown in figure \ref{carpetas}.

Let us now discuss on the concepts of fractal dimensionality and 
discrete scale invariance associated with the SC's used. 
The fractal dimension $d_f$ characterizes the dependence of the mass $M(L)$
(or equivalently, the number of occupied sites of the fractal) 
as a function of the linear size $L$ of the system, so that if we 
now consider an amplification of the system of size $bL$, one has

\begin{equation}
M(bL)= b^{d_{f}} M(L),
\label{massf}
\end {equation}

\noindent which yields the following solution

\begin{equation}
M(L) = B L^{d_{f}} , 
\label{massf2}
\end {equation}

\noindent where $B$ is a constant and $d_{f}$ is a noninteger dimension 
known, after Mandelbrot, as the fractal dimension. By applying 
equation (\ref{massf}) to the carpets $SC_{x}(l,N_{occ},k)$
one gets $d_{f}=ln(N_{occ})/ln(l)$, which leads to $d_{f}=ln(5)/ln(3) \simeq 1.465 $
for the carpets shown in figure \ref{carpetas} with $l=3$ and $N_{occ}=5$.

In general, the factor $b$ in equation (\ref{massf})
could be an arbitrary real number, leading to \textit{continuous} scale invariance.
However, deterministic fractals exhibit \textit{discrete} scale 
invariance (DSI) \cite{refsornosa}, which is a weak kind of scale invariance such 
that $b$ is no longer an arbitrary real number, but it can only take specific 
discrete values of the form $b_n =(b_1)^n$, where $b_1$ is a fundamental scaling 
ratio. Then, for the case of DSI, the solution of equation (\ref{massf}) yields

\begin{equation}
M(L) = L^{d_{f}} F\left( \frac{\log (L)}{\log (b_1)}\right),  
\label{ecdsi2}
\end{equation}

\noindent where $F$ is a periodic function of period one. Notice that for the 
$SC_{x}(l,N_{occ},k)$ one has that $b_1 = l$. The measurement of soft 
oscillations in spatial domain \cite{refsornosa} is a signature of spatial DSI.

Very recently one of us found evidence of discrete scale invariance
in the {\it time} domain by measuring the relaxation of the magnetization
in the Ising model on Sierpinski carpets \cite{we2,we}. Subsequently, it has
been conjectured that physical processes characterized by an observable $O(t)$, 
occurring in fractal media with DSI, 
and that develop a monotonically increasing time-dependent characteristic length $\xi(t)$,
may also exhibit {\it time} DSI \cite{lett}. In fact, by assuming 

\begin{equation}
\xi \propto t^{1/z_{D}}, 
\label{zeta}
\end{equation}

\noindent where $z_{D}$ is a dynamic exponent, it can be shown that $O(t)$ has 
to obey time DSI according to \cite{lett} 
 
\begin{equation}
O(t) = C t^{\gamma /z_{D}} F\left( \phi + \frac{\log (t)}{\log (b_{1}^{z_{D}})} \right),  
\label{ecdsi3}
\end{equation}

\noindent where $C$ and $\phi$ are constants, and $\gamma$ is the relevant exponent
in the expected power-law behavior of the observable $O(t)$. So, 
the conjecture given by equation (\ref{ecdsi3}) implies the existence of 
a logarithmic periodic modulation of time observables characterized by 
a time-scaling ratio $T$ given by

\begin{equation}
T = b_{1}^{z_{D}},
\label{app2}
\end{equation}

\noindent see also equation (\ref{ecdsi2}). It is worth mentioning that
in the case of growth models, equation (\ref{zeta}) can be identified, e.g.,
with the time development of the correlation length along the direction parallel to the 
interface given by $\xi_{\parallel} \propto t^{1/z}$.   

As follows from figure \ref{carpetas}, in this work only fractals
with finite ramification order \cite{frac} are used. A finite ramification implies 
that the fractal structure has weak points, where only a finite number of 
links connect two parts of arbitrary size together. 

\section{The BD model on fractal substrates and definition of its internal structure}

The lattice version of BD is simple to describe: particles fall vertically 
onto the substrate from a random position above the surface. When a particle 
reaches the surface, it sticks on the first site encountered that is a 
nearest-neighbor of an already deposited particle. Due to this constraint
the growth of an interface essentially parallel to the substrate is observed.
For substrates of integer dimension, the BD model can be described by the
Kardar-Parisi-Zhang (KPZ) equation \cite{BS,KPZ}, namely, 

\begin{equation}  
\frac{\partial h({\mathbf x},t)}{\partial t}= \mathcal{F}+\nu_{o} \nabla ^{2}h({\mathbf
x},t)+  \frac {\lambda}{2}[\nabla h({\mathbf x},t)]^{2}+\eta ({\mathbf x},t). 
\label{eq28} 
\end{equation}

\noindent In this equation the nonlinear term represents the 
lateral growth or the appearance
of a driven force, $\nu_{o}$ accounts for the effective surface tension, 
$\mathcal{F}$ is the 
flux of incoming particles, and $\eta ({\mathbf x},t)$ is a Gaussian noise with
zero configurational average. In $d=1$ dimension, equation (\ref{eq28}) can be 
solved exactly and the resulting exponents are  $z = 3/2$, $\alpha = 1/2$ 
and $\beta = 1/3$. For $d=2$ dimensions, one still lacks an analytical solution, 
but according to numerical simulations one has estimations of the exponents given 
by  $\alpha \sim 0.40$  and $\beta \sim 0.24$ \cite{cas}.
  
In order to describe the internal structure of BD aggregates, one considers that 
trees are formed by assuming that any newly deposited particle belongs to the 
same tree as that of the nearest neighbor particle where it is attached \cite{trees}. 
Also, if the deposited particle has more than one nearest neighbor belonging to 
different trees, one of them is selected at random and the particle 
is incorporated into that tree. Relevant exponents (see equations (\ref{Eq.1}),
(\ref{Eq.2}), and (\ref{Eq.3})) have already been determined 
yielding $\nu _{\parallel} =0.60(1)$, $\nu _{\perp}=0.40(1)$, 
and $\tau=1.40(1)$ in $d=1$ dimension, and $\nu _{\parallel} =0.45(1)$, 
$\nu _{\perp}=0.29(1)$, and $\tau=1.57(1)$ in $d=2$ dimensions (see, e.g., 
\cite{trees} and references therein).  

On the other hand, for BD growth on fractal substrates, particles are deposited 
randomly on {\bf occupied sites of the fractal only} by following the rules 
described above for the case of integer-dimensional substrates.   
Also the building procedure employed to determine the trees used for the 
description of the internal structure of the whole aggregate is independent of 
the substrate.

For the purpose of a numerical simulation, the Monte Carlo time step (mcs) involves
the deposition of $L^{d_{f}}$ particles.

\section{Application of the Family-Vicsek phenomenological dynamic scaling approach} 

\begin{figure}[h]
\includegraphics[width=10.0cm,clip=true,angle=270]{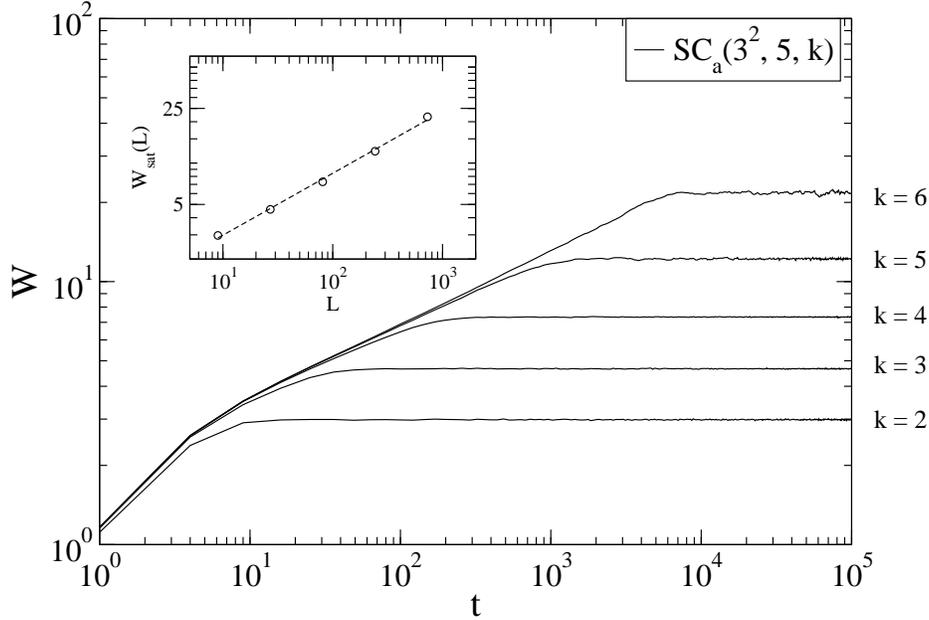}
\caption{Log-log plots of the interface width ($W$) versus time for the 
BD model on $SC_{a}(3^{2},5,k)$ with $k=2-6$. The inset  shows a log-log plot 
of $W_{sat}(L)$ versus $L$. The dotted line with slope $0.47$ corresponds to the best fit  
of the data.}
\label{escaleo1}
\end{figure}

Figure \ref{escaleo1} shows log-log plots of $W$ versus $t$ obtained for 
the BD model on fractal substrates obtained by using the generating 
cell $SC_{a}(3,5,1)$. For short times, say $t < 3$ mcs, the random growth 
of the interface is observed because the random deposition (RD) process 
dominates. At this stage, correlations have not been developed yet and 
one has that, according to equation (\ref{FV}), 
$W(t)$ $\propto t^{\beta _{RD}}$, with $\beta _{RD} = 1/2$, holds.  
During an intermediate time regime, say $3$ mcs $ < t < t_{c}$, 
correlations develop since the BD process now prevails, leading 
to the typical growth regime $W(t)$ $\propto t^{\beta_{BD}}$,
see also equation (\ref{FV}).  
At a later stage, for $t > t_{c}$, correlations can no longer develop 
due to the geometrical constraint of the lattice size, and
the saturation regime ($W_{sat}(L) \propto L^{\alpha_{BD}}$) is observed,
as expected from equation (\ref{FV}). Here, $t_{c}  \propto L^{z}$ is the 
crossover time between the growing and the saturation regimes of 
the interface width. 

\begin{figure}[h]
\includegraphics[width=10.0cm,clip=true,angle=270]{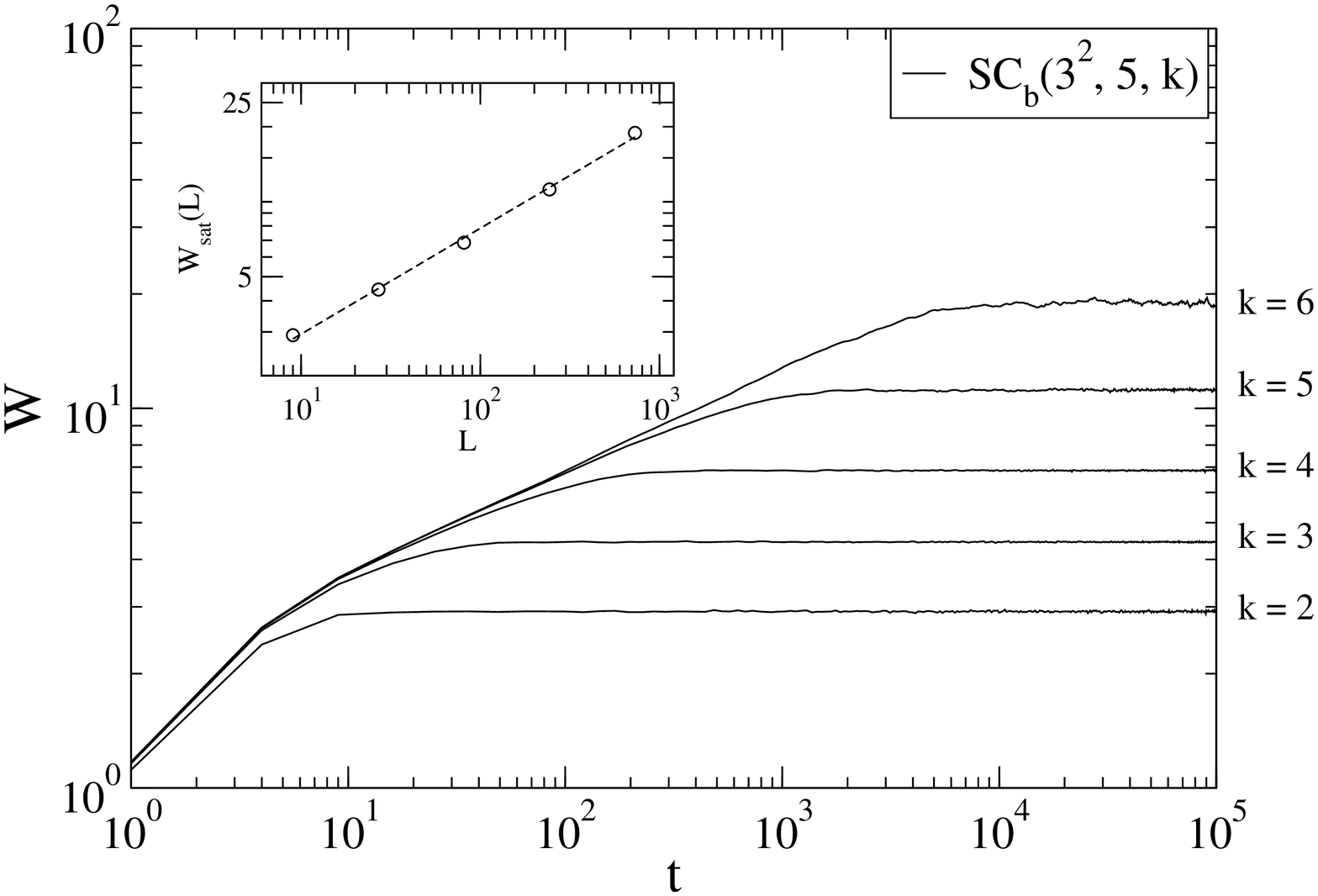}
\caption{Log-log plots of the interface width ($W$) versus time for the 
BD model on $SC_{b}(3^{2},5,k)$ with $k=2-6$. The inset  shows a log-log plot 
of $W_{sat}(L)$ versus $L$. The dashed line with slope 0.44 corresponds to the best fit  
of the data.}
\label{escaleo2}
\end{figure}

\begin{figure}[h]
\includegraphics[width=10.0cm,clip=true,angle=270]{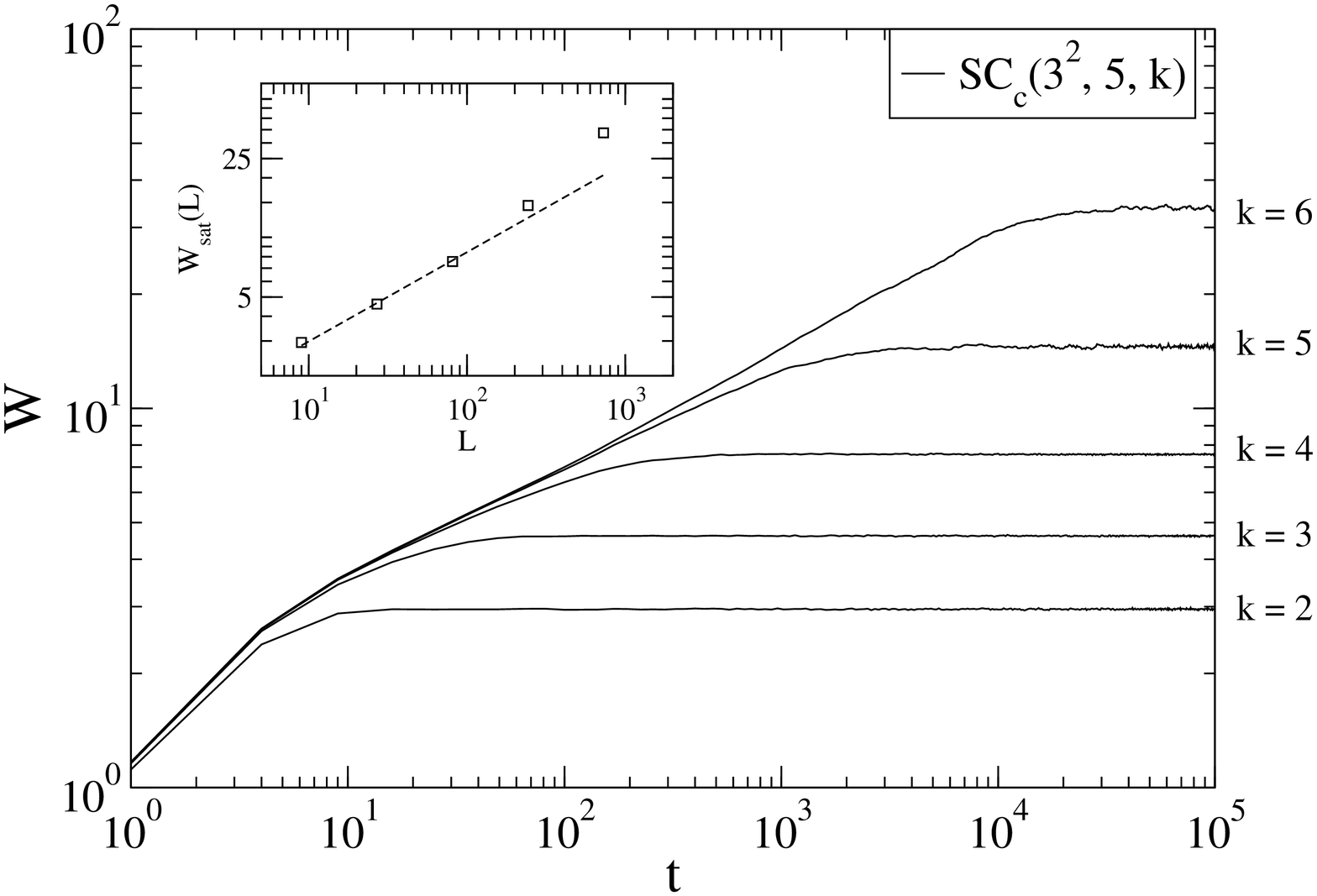}
\caption{Log-log plots of the interface width ($W$) versus time for the 
BD model on $SC_{c}(3^{2},5,k)$ with $k=2-6$. The inset  shows a log-log plot 
of $W_{sat}(L)$ versus $L$. The dashed line with slope 0.48 have been drawn 
for the sake of comparison.}
\label{escaleo3}
\end{figure}

In order to obtain the roughness, growth, and dynamic exponents, we followed the 
standard procedure \cite{BS}, e.g., the inset in figure \ref{escaleo1} shows 
a log-log plot of $W_{sat}(L)$ versus $L$. Here, a slight but noticeable systematic 
upward deviation of the data is observed. However, an effective exponent ($\alpha$) 
can be determined and the best fit of the data yields $\alpha = 0.47 \pm 0.04$. 
Also, we determined $\beta = 0.28 \pm 0.03$ and $z = 1.56 \pm 0.05$ for the BD model on 
the fractal substrate obtained by using the $SC_{a}(3,5,1)$ generating cell. 

Figures \ref{escaleo2} and \ref{escaleo3} show log-log plots of $W$
versus $t$ obtained for the BD model on fractal substrates built up by using
the generating cells $SC_{b}(3,5,1)$ and $SC_{c}(3,5,1)$, respectively.
Again, the standard procedure was attempted in order to determine
the roughness, growth, and dynamic exponents. Accordingly, the insets in
figures \ref{escaleo2} and \ref{escaleo3} show log-log plots of
$W_{sat}(L)$ versus $L$.
For the case of the $SC_{b}(3,5,1)$, again, a noticeable systematic
upward deviation of the data is observed, but an effective exponent ($\alpha$) can be
determined, yielding $\alpha = 0.44 \pm 0.04$. Also, we determined $\beta = 0.26 \pm 0.03$
and $z = 1.55 \pm 0.05$. On the other hand, for the $SC_{c}(3,5,1)$ we
observed strong systematic deviations of the data that prevent the evaluation of the exponents.
This results strongly suggest that in the case of aggregates grown on
fractals generated by $SC_{c}(3,5,1)$ either the finite-size effect are very important and 
one has to introduce important scaling corrections to the standard Family-Vicsek approach, 
or (instead of we are not be able to identify two 
competing processes leading to the formation of the aggregate) eventually the onset 
of a crossover effect is present. The understanding/clarification of which one of the two possibilities may be true is a very difficult task, which deserves further work, and is beyond the scope of the present paper.

Anyway, the obtained effective exponents are almost the same (within error bars) 
for both the $SC_{a}(3,5,1)$ and the $SC_{b}(3,5,1)$, namely
$\alpha \approx 0.46 $,  $\beta \approx 0.27 $, and $z \approx 1.55$. It is
worth mentioning that these exponents nicely interpolates between the exact values 
corresponding to $d = 1$ and the best estimates reported for $d = 2$, namely
$0.24 (d = 2) < \beta  \approx 0.27 < 1/3 (d = 1)$, 
$0.40 (d = 2) < \alpha \approx 0.46 < 1/2 (d = 1)$, and
$1.67 (d = 2) > z      \approx 1.55 > 3/2 (d = 1)$.     
 
\section{Characterization of the internal structure of the aggregates} 

\begin{figure}[h]
\includegraphics[width=10.0cm,clip=true,angle=270]{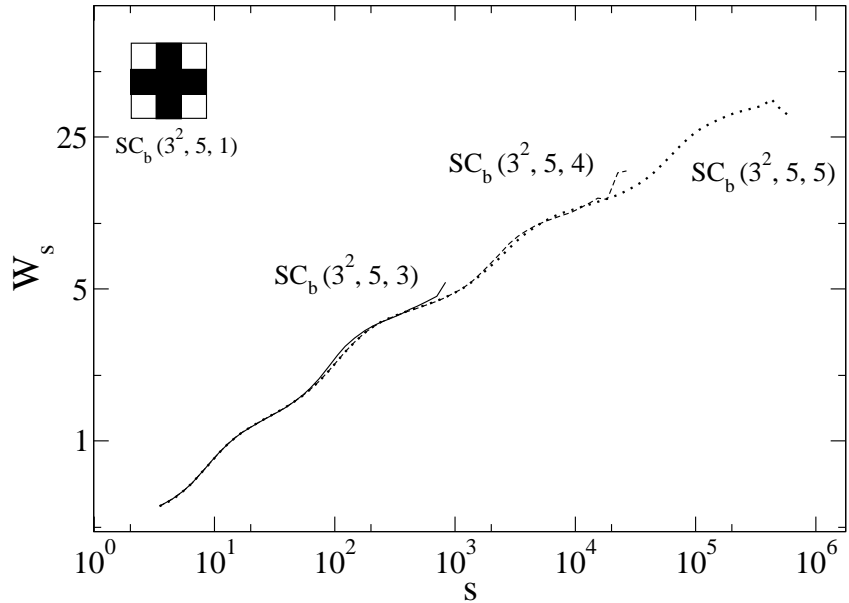}
\caption{Log-log plots of rms width of the trees as a function of the
tree size  $s$ for the BD model on $SC_{b}(3,5,k)$ substrates ($k=3,4,5$).}
\label{ancho1}
\end{figure}

\begin{figure}[h]
\includegraphics[width=10.0cm,clip=true,angle=270]{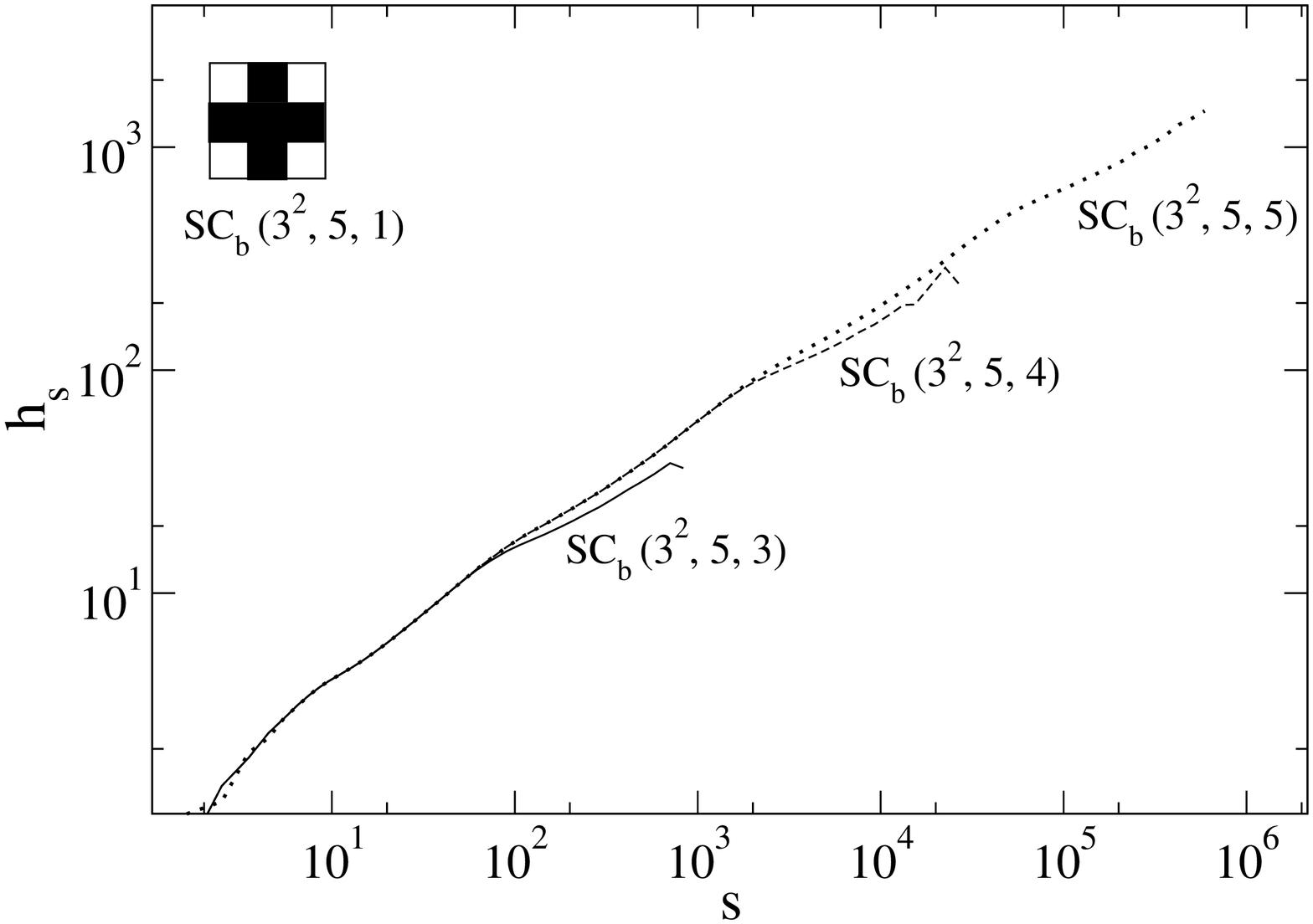}
\caption{Log-log plots of rms height of the trees as a function of the
tree size  $s$ for the BD model on $SC_{b}(3,5,k)$ substrates ($k=3,4,5$).}
\label{altura1}
\end{figure}

The dependence of the rms width and rms height of the trees forming the aggregates 
on the tree size \(s\), for the cases of the $SC_{b}(3,5,k)$ substrates ($k=3,4,5$),
is shown in log-log plots in figures \ref{ancho1} and \ref{altura1}, respectively. 
For the growth of the BD model on nonfractal substrates it is known that the
power laws given by equations (\ref{Eq.1}) and  (\ref{Eq.2}) hold \cite{trees}, 
but in the case of fractal media, one also clearly observes soft oscillations 
with a logarithmic period, which modulate the power laws. 
Figures \ref{ancho1} and \ref{altura1} also show 
a direct relationship between the number of observed oscillations  
and the generation ($k$) of the Sierpinski carpets used as substrates.

\begin{figure}[h]
\includegraphics[width=10.0cm,clip=true,angle=270]{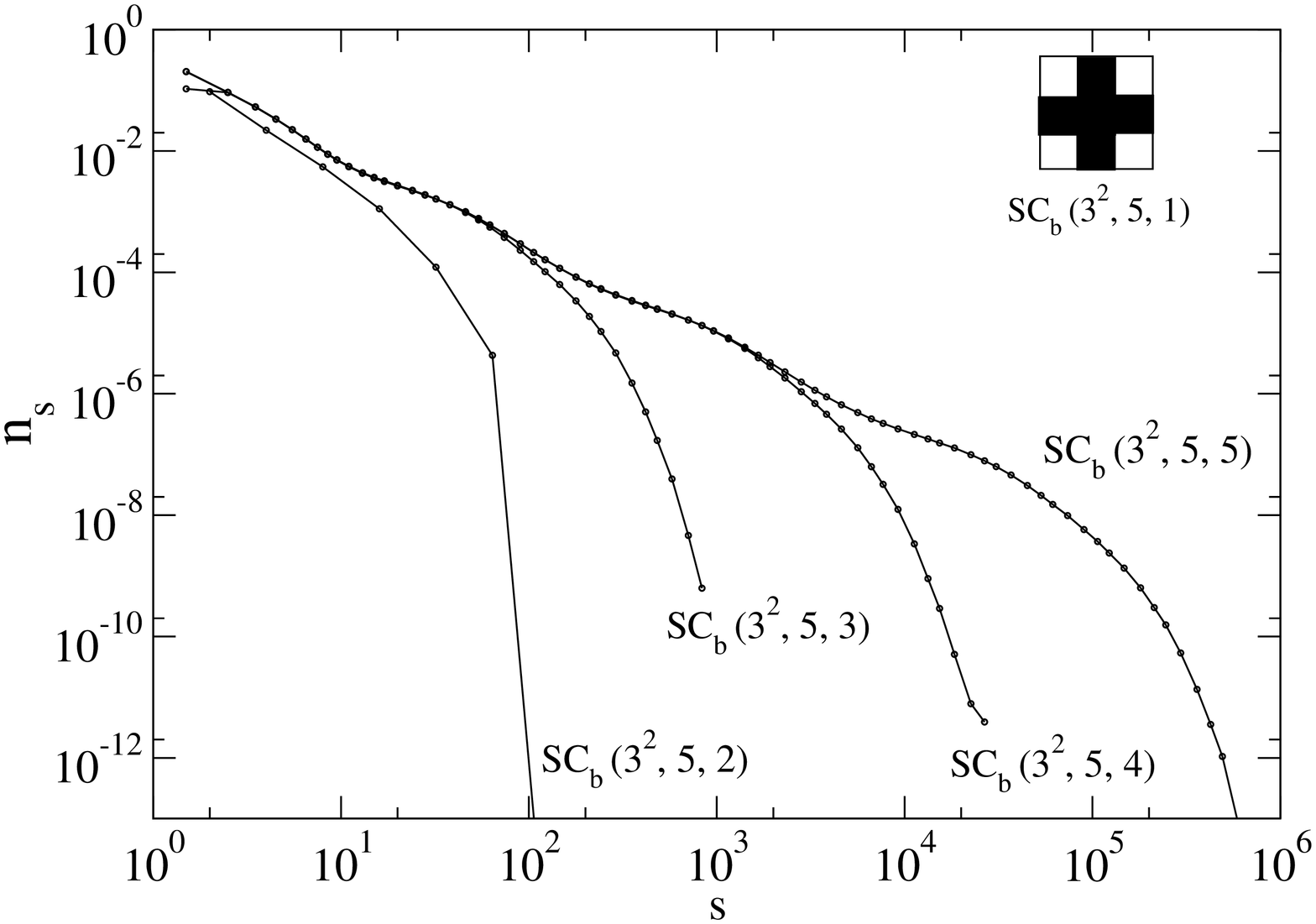}
\caption{Log-log plots of the tree size distribution corresponding to the
BD model on $SC_{b}(3,5,k)$ substrates ($k=2,3,4,5$)}
\label{masa1}
\end{figure}

Figure \ref{masa1} shows log-log plots of the tree-size distribution functions 
corresponding to the BD model grown on $SC_{b}(3,5,k)$ substrates ($k=2,3,4,5$). 
Again, for the BD model on nonfractal substrates the power law given by
equation (\ref{Eq.3}) holds, but when a fractal carpet is used as substrate, a soft 
oscillation with a logarithmic period modulates the power-law behavior. 
Also, there is a one to one relationship between the number of observed
oscillations and the number of generations of the carpet.

\begin{figure}[h]
\includegraphics[width=10.0cm,clip=true,angle=270]{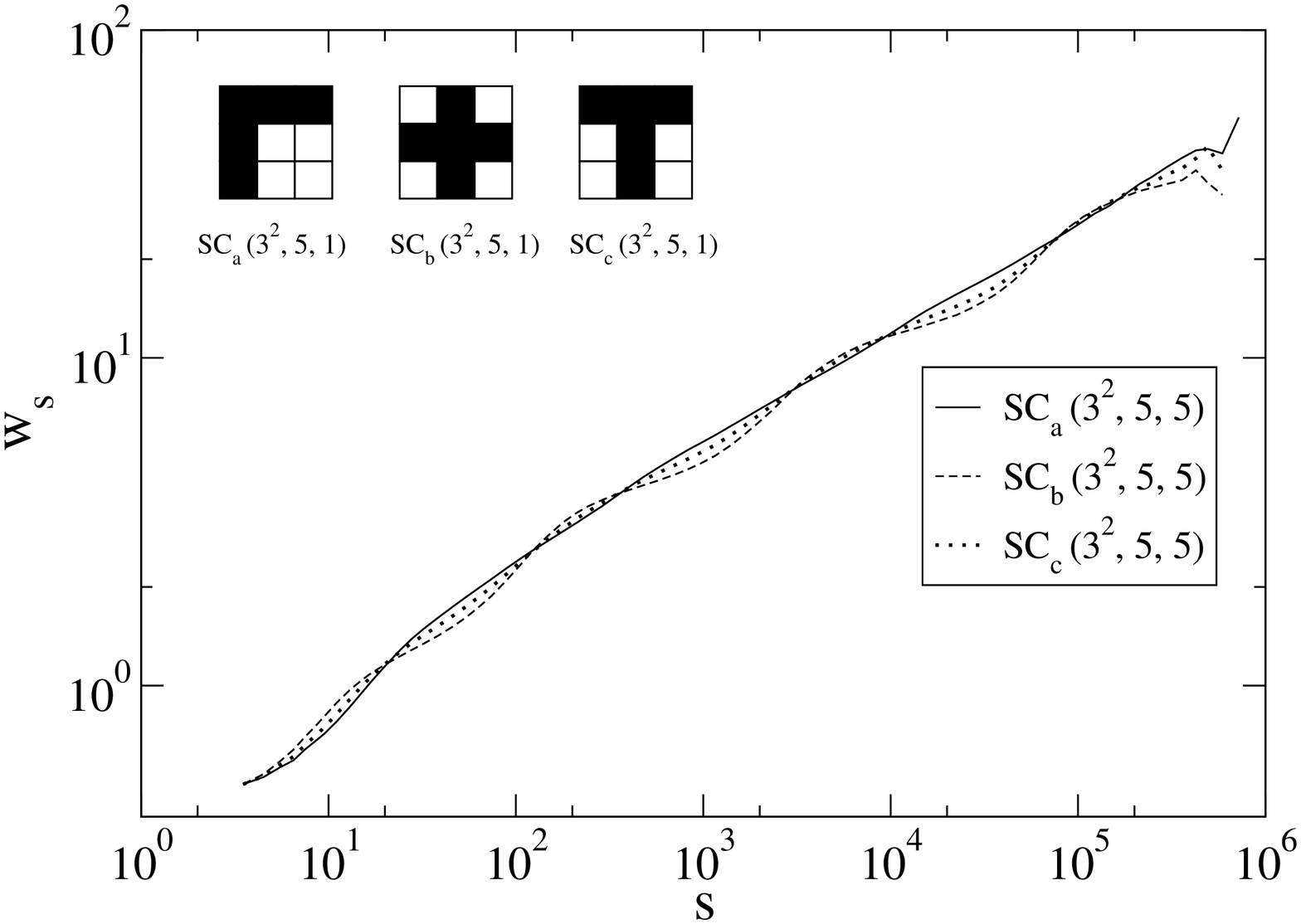}
\caption{Log-log plots of rms width of the trees as a function of the
tree size  $s$ for the BD model on $SC_{x}(3,5,5)$ substrates ($x=a,b,c$).}
\label{ancho2}
\end{figure}

\begin{figure}[h]
\includegraphics[width=10.0cm,clip=true,angle=270]{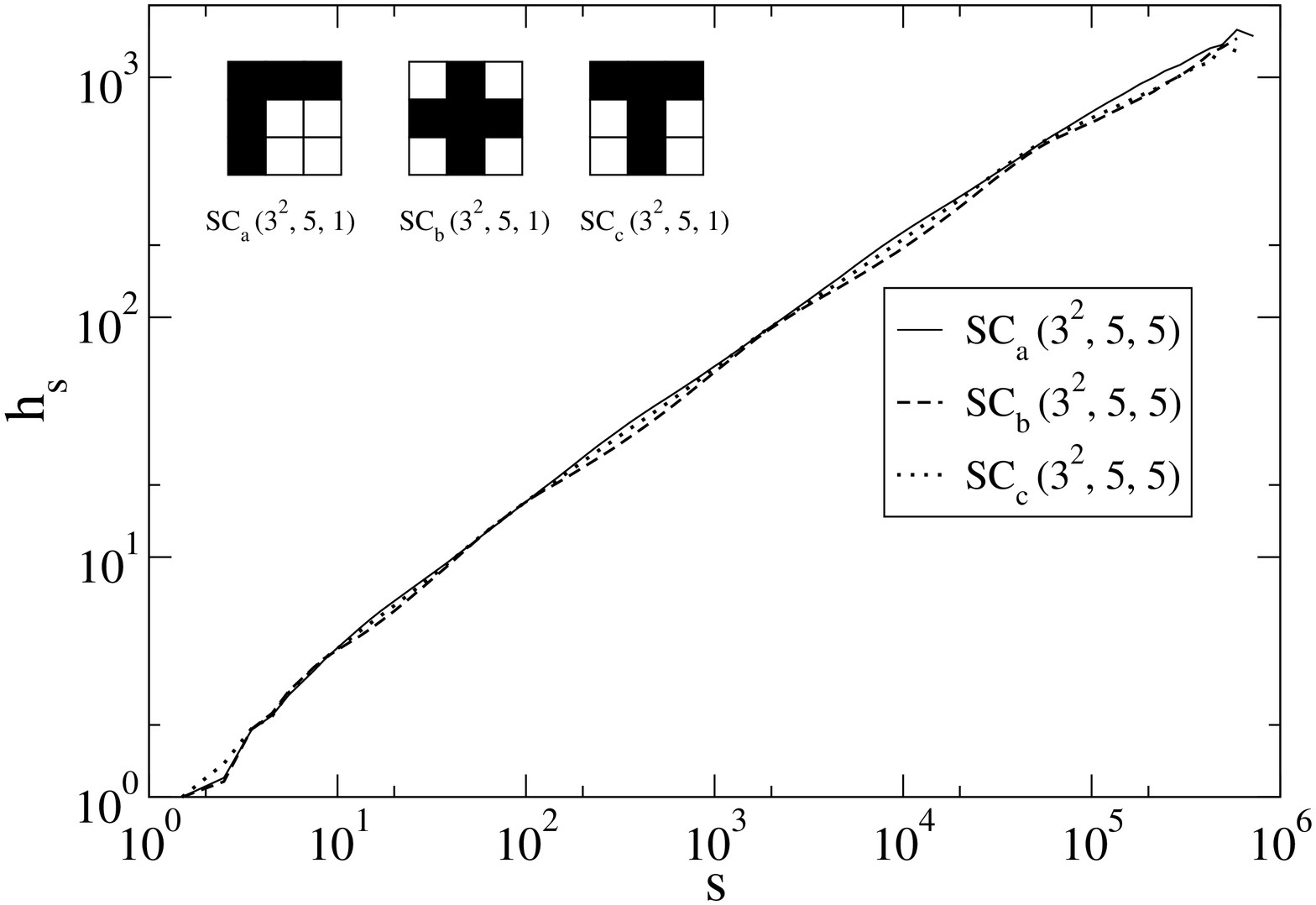}
\caption{Log-log plots of rms height of the trees as a function of the
tree size  $s$ for the BD model on $SC_{x}(3,5,5)$ substrates ($x=a,b,c$).}
\label{altura2}
\end{figure}

\begin{figure}[h]
\includegraphics[width=10.0cm,clip=true,angle=270]{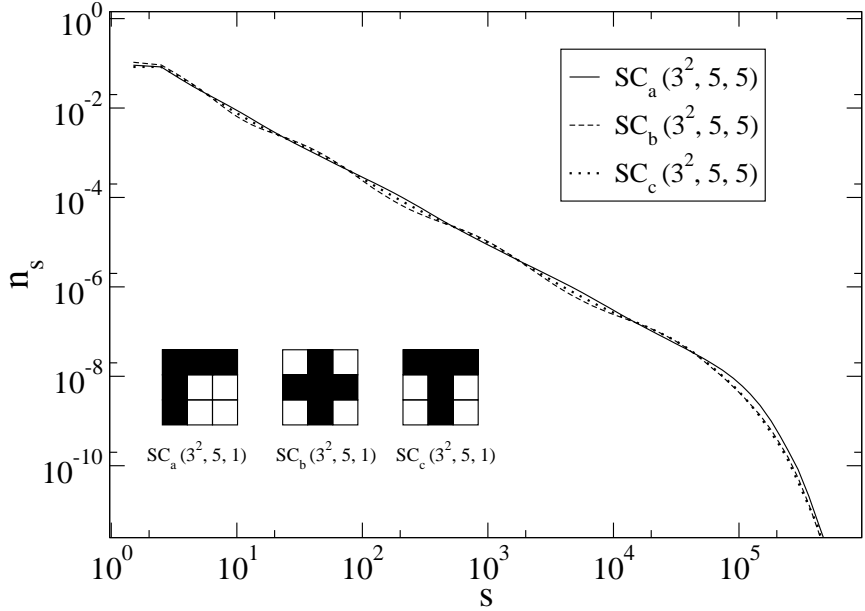}
\caption{Log-log plots of the tree size distribution corresponding to the
BD model on $SC_{x}(3,5,5)$ substrates ($x=a,b,c$).}
\label{masa2}
\end{figure}

\begin{figure}[h]
\includegraphics[width=10.0cm,clip=true,angle=270]{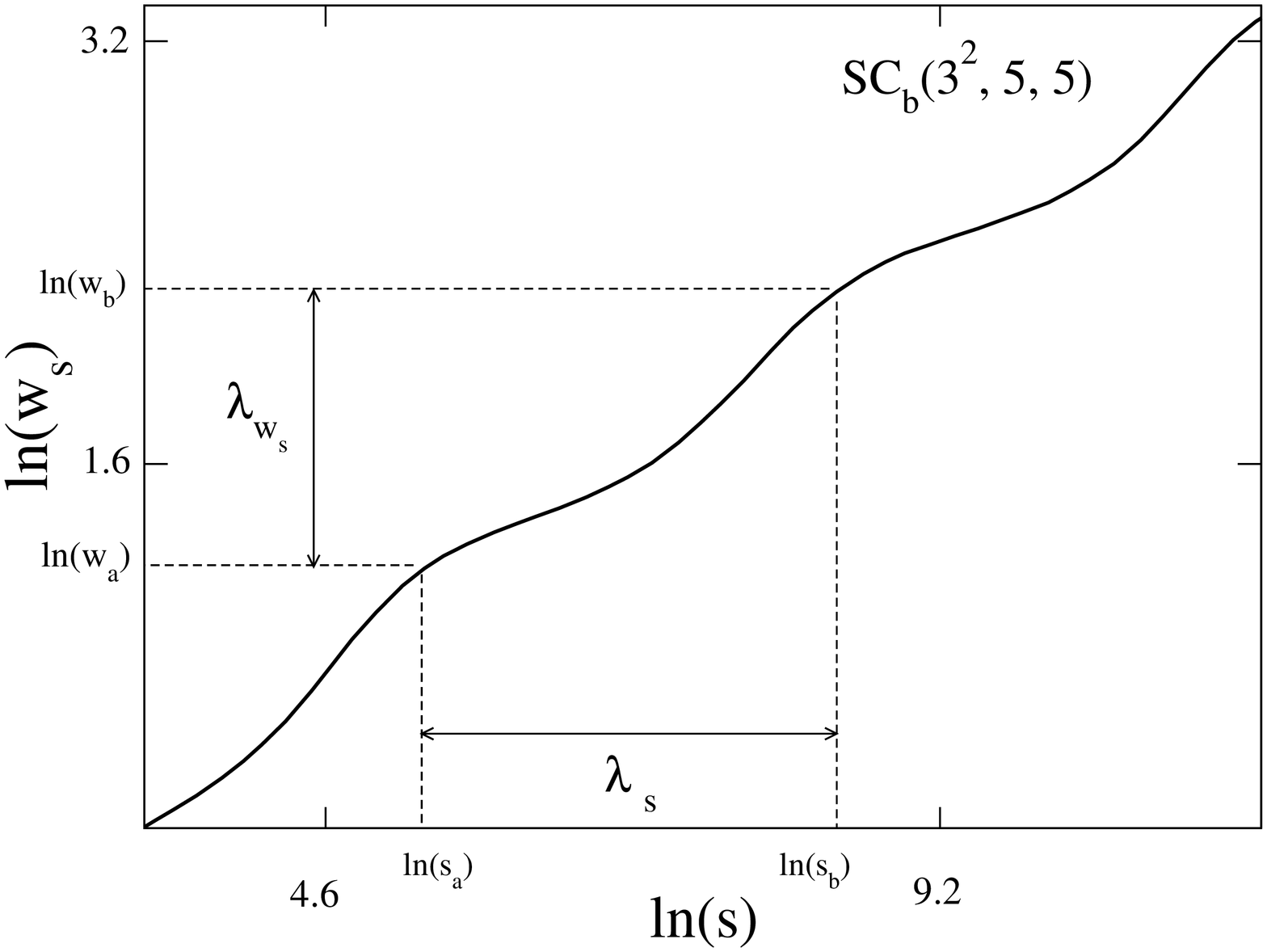}
\caption{Schematic view of the dependence of the rms width of the trees ($w_{s}$) 
versus the tree size ($s$), showing both the logarithmic period of the  
size of the trees ($\lambda_{s}$), and the logarithmic period of the rms 
width ($\lambda_{w_s}$).}
\label{periodo}
\end{figure}

After a systematic study, we found that the rms width, the rms height, and the
tree-size distribution for BD aggregates grown on different
deterministic fractal substrates, generated by $l=3$ and $N_{occ}=5$, exhibit a quite
similar behavior, namely, soft oscillations with a logarithmic period, which modulate
the expected power laws. Figures \ref{ancho2}, \ref{altura2}, and \ref{masa2} show
log-log plots of the rms width, the rms height, and the tree-size distribution,
obtained for these fractal substrates, respectively.
Table I summarizes the exponents obtained by fitting the modulated power laws describing the
properties of the internal structure of the aggregates. Also, data
corresponding to the exponent $z$ obtained
by means of the Family-Vicsek dynamic scaling approach have been
included for the sake of comparison. The values of the 
exponents $Y \equiv \nu_{\parallel}$, $\nu_{\perp}$, and $\tau$, have been obtained 
by fitting the data with the function

\begin{equation}  
\frac{d \ln(X)}{d \ln(s)} = Y + \sum_{n=1}^{N} b_n cos(n \, \omega \ln(s) +\xi _n), 
\label{ajuste.new} 
\end{equation}

where $X \equiv h_s$, $w_s$, and $n_s$, respectively \cite{andrade}. 
Here, $b_n$ and $\xi _n$ are constants. When the fit is performed by taking 
$N > 1$, the presence of higher harmonics to the fundamental frequency is considered. 
So, the values of the exponents obtained by using $N=1$ could be improved \cite{andrade}. 
However, when we considered higher harmonics, inspite of 
the fact that typically we have $b_2/b_1 \sim 0.2$, the values of the
exponents obtained by fitting the data with $N=1$, and $2$ are almost
indistinguishable (within errors bars).

%\vspace{0.3cm}
\begin{table}[tbp]
\label{TableI}
\caption{List of exponents describing the properties of the
internal structure of ballistic aggregates on different fractals.
The exponent $z$ ($5^{\text th}$ column) is obtained by using equation 
(\ref{Eq.8}) and the values of $\nu_{\perp }$ and $ \nu_{\parallel} $ are listed
in the $2^{\text nd}$ and $3^{\text rd}$ columns, respectively. Also, the exponent
$z$ ($6^{\text th}$ column) is evaluated by using data obtained by applying the 
Family-Vicsek dynamic scaling (equation (\ref{FV})).}   
\vspace{1.0cm}
{\centering \begin{tabular}{|c|c|c|c|c|c|c|c|c|c|}
\hline 
 Generating cell         &
\( \nu_{\perp }      \)  &
\( \nu_{\parallel }  \)  &
\( \tau              \)  &
$   z = \nu_{\parallel }/ \nu_{\perp }   $   &
$   z = \alpha/\beta $   \\
\hline 
\hline 
$SC_{a}(3,5,1)$&
$0.34(1)$&
$0.54(1)$&
$1.48(1)$&
$1.59(6)$&
$1.56(5)$\\
\hline 
$SC_{b}(3,5,1)$&
$0.34(1)$&
$0.55(1)$&
$1.48(1)$&
$1.60(6)$&
$1.55(5)$\\
\hline 
$SC_{c}(3,5,1)$&
$0.35(1)$&
$0.56(1)$&
$1.48(5)$&
$1.61(6)$&
$--$\\
\hline 
\hline 
\end{tabular}\par}
\end{table}
\vspace{1.0cm}

\begin{figure}[h]
\includegraphics[width=10.0cm,clip=true,angle=270]{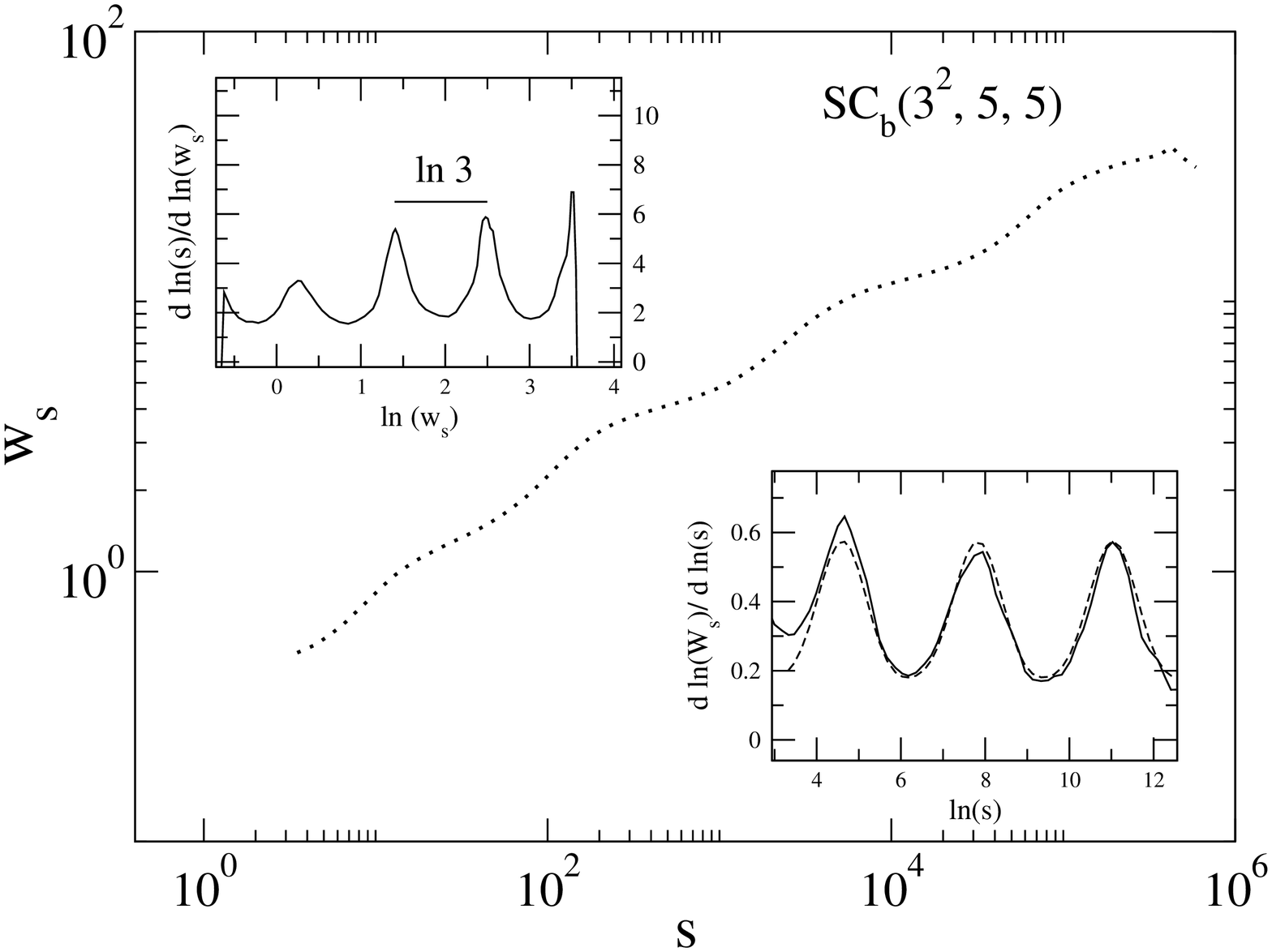}
\caption{Log-log plots of the rms width of the trees as a function of the tree 
size \(s\) as obtained for ballistic aggregates on $SC_{b}(3^{2},5,5)$ substrates. 
The derivative of $\ln(w_{s})$ with respect to $\ln (s)$ is shown in the inset placed 
at the lower-right hand side, while the derivative of $\ln(s)$ with respect to 
$\ln (w_{s})$ is shown in the inset placed on the upper-left hand side. The dashed line in 
the inset placed at the lower-right hand side correspond to the best fit of the data 
using Eq. \ref{ajuste.new}.}
\label{ancho3}
\end{figure}

In order to rationalize our findings, let us discuss the soft oscillations observed 
in the rms width mentioned above. Here, we can identify two logarithmic periods,
one for the size of the trees ($\lambda_{s}$) and the other for the rms 
width ($\lambda_{w_s}$), as is schematically shown in figure \ref{periodo}. 
These periods become evident by calculating the derivative of $\ln(w_{s})$ with 
respect to $\ln (s)$, and the derivative of $\ln(s)$ with respect to $\ln (w_{s})$, 
as shown in figure \ref{ancho3} for the case of the $SC_{b}(3,5,5)$ substrate.  
Also, it is worth mentioning that by following the same procedure, we can 
identify one logarithmic period for the rms height ($\lambda_{h_s}$) and 
another one for the size distribution ($\lambda_{n_{s}}$). Furthermore, as mentioned 
above, there is a one to one relationship between the number of observed oscillations 
and the generations of the carpets. So, as all measurements share the same range 
of values of $s$, the period of the oscillation observed for the size of 
the trees ($\lambda_{s}$), when $k$ tends towards infinity, has to be the same for all 
observables, namely, the rms width, the rms height, and the tree-size distribution.

Now, based on both the already discussed  results and equation (\ref{Eq.2}), 
we conjecture that for the case of a fractal substrate and in the limit 
$k \rightarrow \infty $, the derivative of $\ln(w_{s})$ with respect to 
$\ln (s)$ is a periodic function (with a logarithmic period) that 
modulates a constant value. So, the conjecture can be written as

\begin{equation}  
\frac{\partial \ln(w_s)}{\partial \ln(s)} = \nu_{\perp } + \zeta (\ln(s)), 
\label{eqa001} 
\end{equation}

\noindent where $\zeta (x)$ is a periodic function. By integration of
equation (\ref{eqa001}) it follows that  

\begin{equation}  
w_s = A s^{\nu_{\perp }} \exp^{\zeta^{*} (\ln(s))} ,  
\label{a002} 
\end{equation}

\noindent where $A$ is a constant and $\zeta (x) ^{*}$ is a periodic function 
such that $\partial \zeta^{*} (\ln(s)) / \partial \ln(s) = \zeta (\ln(s))$.   

Let us now take two tree sizes, $s_a$ and $s_b$, separated by a period $\lambda_{s}$ 
(on a logarithmic scale), as shown in the horizontal axis of figure \ref{periodo}. Also,  
their corresponding rms widths, called $w_a$ and $w_b$, respectively, are separated 
on a logarithmic scale by a period $\lambda_{w_{s}}$ (see figure \ref{periodo}). 
Then, by assuming that equation (\ref{a002}) holds, it can be found that 

\begin{equation}  
\lambda_{w_s} = \nu_{\perp }  \lambda_{s}.
\label{a003} 
\end{equation}
 
The existence of a logarithmic period  in the rms width of the trees ($\lambda_{w_s}$)
can be understood due to the fact that each tree can only spread over the fractal and, 
consequently, the rms width of the trees is constrained by the geometrical features of 
the underlying structure. In the case of a deterministic Siernspisky 
carpet, that structure is constructed by the iteration of a generating cell,
so that the topological details of the generating cell are present in all the sample. 
In our simulations, a generating cell of side $l=3$ leads to the occurrence
of discrete scale invariance, such that the underlying structures are self-similar only at 
scales of size $l^n$, where $n$ is an integer. Based on theses concepts, it is expected 
that $\lambda_{w_s} = \ln(3)$. In order to check this statement, in the inset shown 
on the left-hand side of figure \ref{ancho3} we have drawn a line of size $\ln(3)$, 
which nicely fits the logarithmic period of the oscillation of the rms width. 
This result has also been carefully checked for all the studied cases by measuring 
the peak-to-peak distance of the modulating oscillations. This finding leads us to 
conjecture that, for aggregate grown on deterministic fractal substrates generated 
by a generating cell $SC_{x}(l,N_{occ},1)$, the logarithmic 
period  in the rms  width of the trees should obey the following relationship  
 
\begin{equation}  
\lambda_{w_s} = \ln (l).
\label{a004} 
\end{equation}

Equation (\ref{a004}) reflects the fact that the size of the trees in the 
direction parallel to the substrate exhibits spatial discrete scale invariance 
with the same fundamental scaling ratio as that of the underlying fractal, namely, 
$b_{1} = l$ (see also equation (\ref{ecdsi2}) ). 

\begin{figure}[h]
\includegraphics[width=10.0cm,clip=true,angle=270]{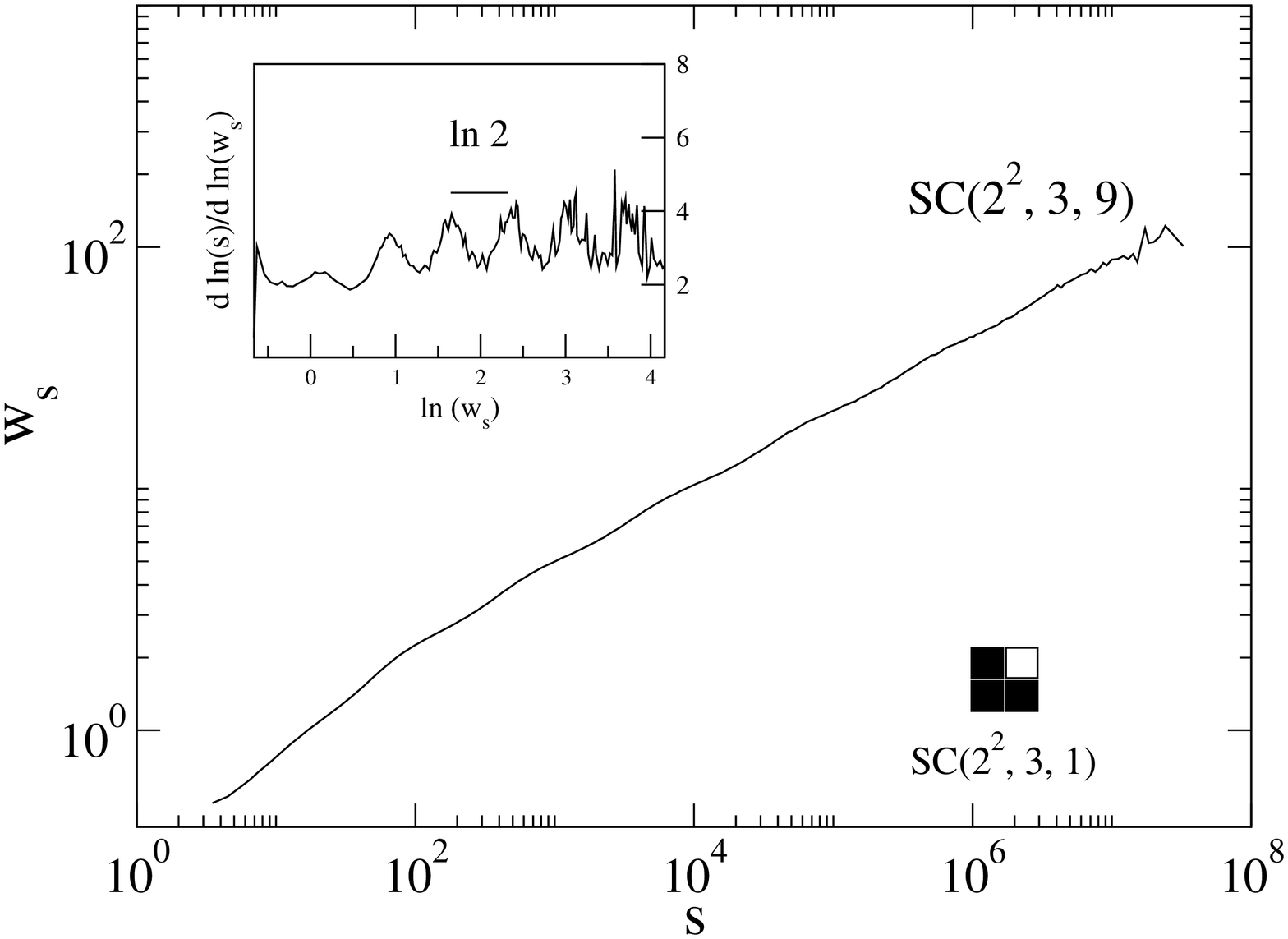}
\caption{Log-log plots of the rms width as a function of the tree 
size \(s\) as obtained for the $SC_{b}(2,3,9)$ substrate. The derivative of 
$\ln(s)$ with respect to $\ln (w_{s})$ is shown in the inset placed on the left.}
\label{ancho2.2}
\end{figure}

In order to check this conjecture, figure \ref{ancho2.2} shows a log-log plot of 
the rms width as a function of the tree size $s$ for the case of an $SC_{b}(2,3,8)$ 
substrate. In this case, the size of the generating cell is $l=2$, so according to 
the conjecture we expect $\lambda_{w_s} = \ln(2)$. The inset placed on the left 
of figure \ref{ancho2.2} shows the derivative of $\ln(s)$ with respect to $\ln (w_{s})$, 
and a line of size $\ln(2)$, which nicely fits the value expected for 
$\lambda_{w_s}$, has been drawn.

\begin{figure}[h]
\includegraphics[width=10.0cm,clip=true,angle=270]{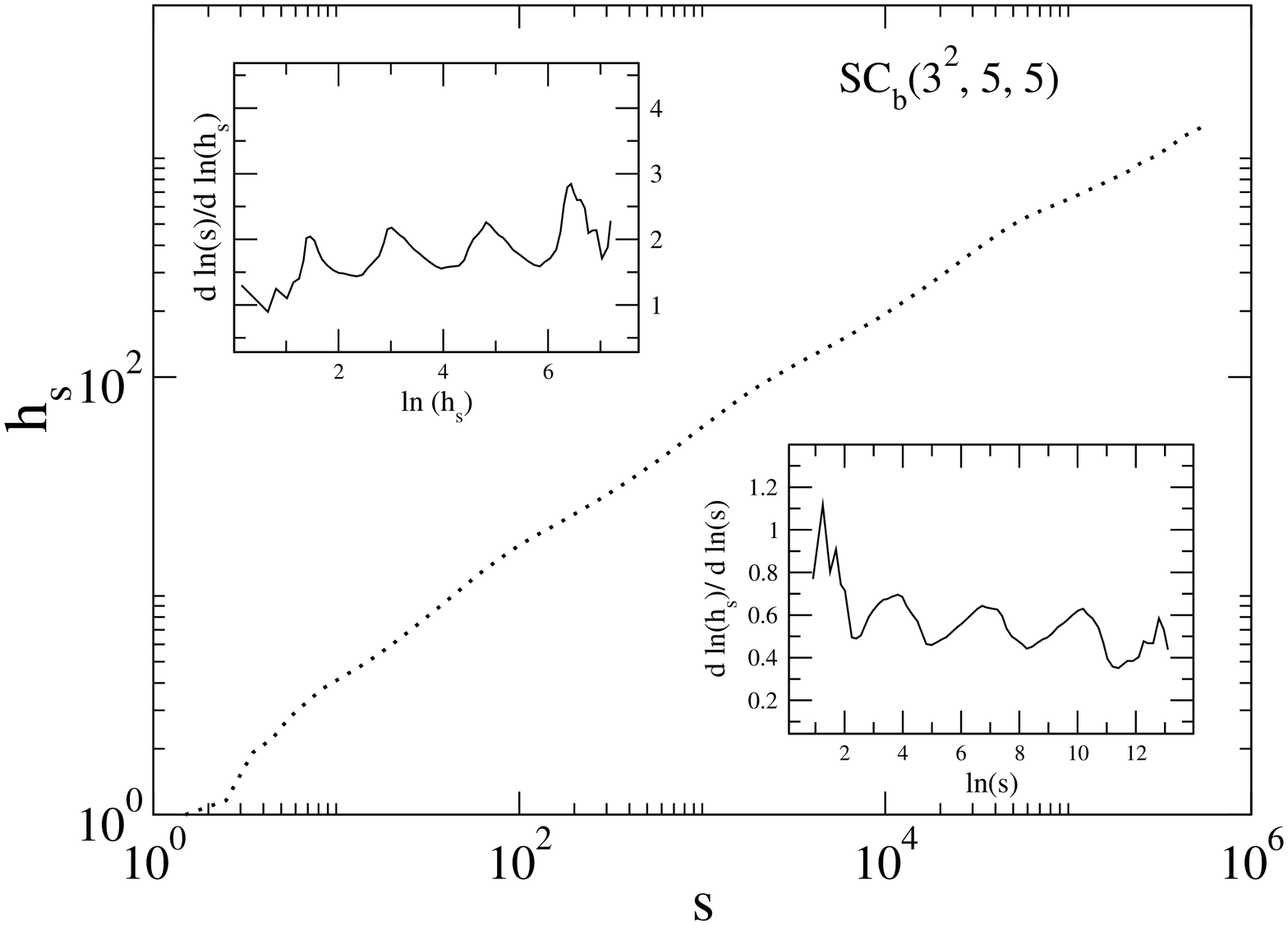}
\caption{Log-log plots of the rms height as a function of the tree 
size \(s\) as obtained for the $SC_{b}(3,5,5)$ substrate. The derivative of 
$ln(h_{s})$ with respect to $ln (s)$ is shown in the inset placed on the right, 
and the derivative of the $\ln(s)$ with respect to $\ln(h_{s})$ is shown in the 
inset placed on the left.}
\label{altura3}
\end{figure}

In order to understand the oscillatory modulation of the power laws observed in the 
rms height of the trees, we followed the same method already discussed for the case 
of the rms width. Figure \ref{altura3} shows the rms height as a function of the tree 
size \(s\), as obtained for the $SC_{b}(3,5,5)$ substrate. Also, both the derivative 
of $\ln(h_{s})$ with respect to $\ln (s)$ and the derivative of $\ln(s)$ with 
respect to $\ln (h_{s})$ are shown in the insets placed on the lower right-hand and the 
upper left-hand sides of figure \ref{altura3}, respectively

So, based on equation (\ref{Eq.1}) and the obtained results, we now conjecture that

\begin{equation}  
\frac{\partial \ln(h_s)}{\partial \ln(s)} = \nu_{\parallel } + \eta (\ln(s)), 
\label{eq b001} 
\end{equation}

\noindent where $\eta (x)$ is a periodic function. 
Then, after integration one gets 

\begin{equation}  
h_s = B s^{\nu_{\parallel }} \exp^{\eta^{*} (\ln(s))} ,   
\label{b002} 
\end{equation}

\noindent where $B$ is a constant and $\eta (x) ^{*}$ is a periodic function 
such that $ \partial \eta^{*} (\ln(s)) / \partial \ln(s) = \eta (\ln(s))$.   

By using  equation (\ref{b002}) and following the method already applied  
to the case of the rms width, it can also be found that 

\begin{equation}  
\lambda_{h_s} = \nu_{\parallel }  \lambda_{s}.
\label{b003} 
\end{equation}

Then, by using equations  (\ref{a003}), (\ref{a004}), and (\ref{b003}) one obtains

\begin{equation}  
\lambda_{h_s} = \frac {\nu_{\parallel }}{\nu_{\perp }}  \ln(3) = z \,  \ln(3) =
 z \,  \ln(l) =  z \, \lambda_{ws} ,
\label{b004} 
\end{equation}

\noindent where equation (\ref{Eq.8}) has also been used. Let us now recall that, 
according to our conjecture given by equation (\ref{ecdsi3}), it should be expected that  
the rms high of the trees, which is an observable that evolves along the growing (time) 
direction, will be coupled to the space discrete scale invariance of the underlying 
fractal, then exhibiting {\bf time} discrete scale invariance.
In fact, the coupling is nicely reflected by equation (\ref{b004}), which is
the realization of equation (\ref{app2}), with $T = \exp(\lambda_{h_s})$ being
the fundamental time scaling ratio.

\begin{figure}[h]
\includegraphics[width=10.0cm,clip=true,angle=270]{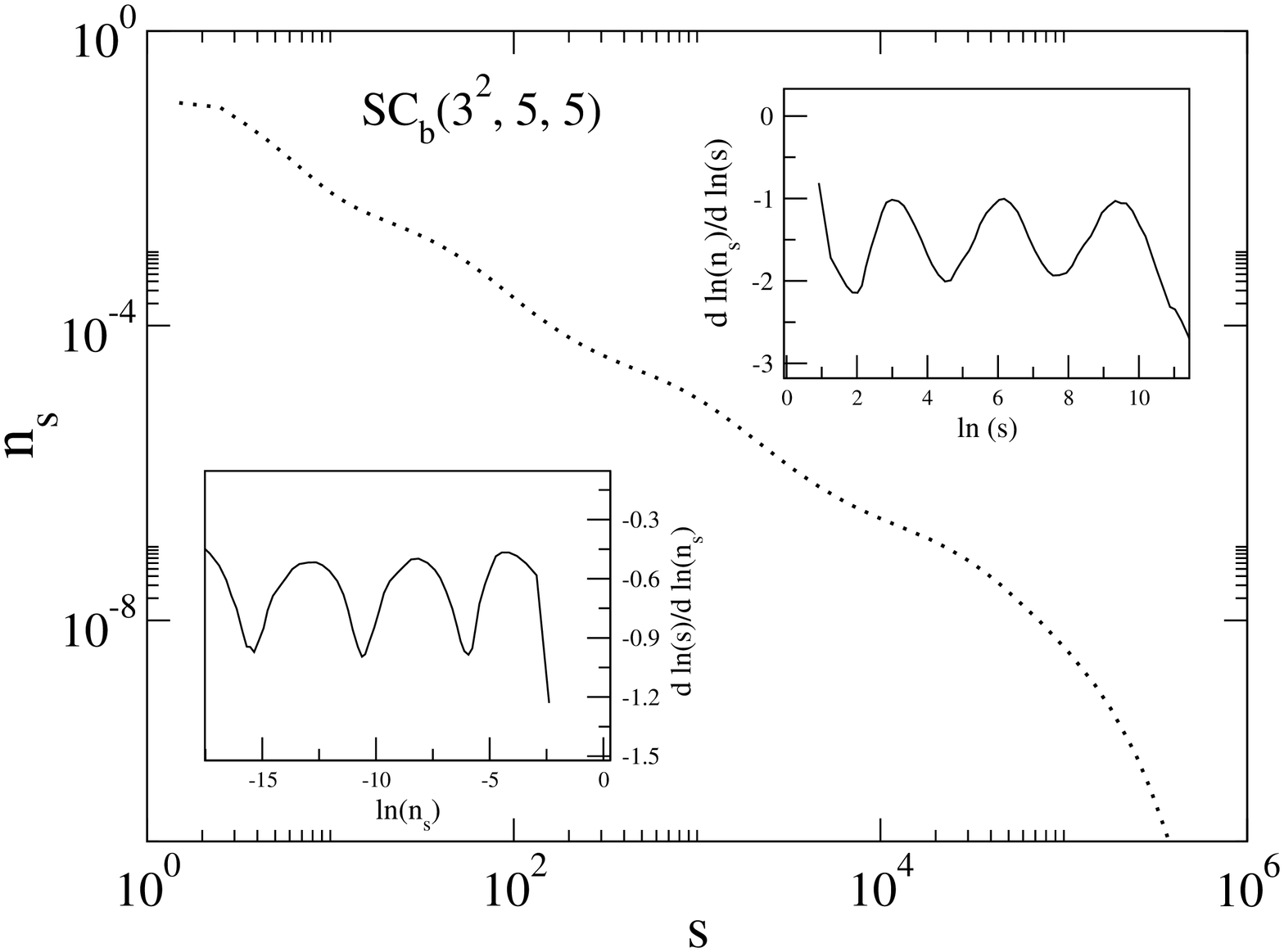}
\caption{Log-log plot of the tree size distribution as obtained for
the $SC_{b}(3,5,5)$ substrate. The derivative of $\ln(n_{s})$ with respect 
to $\ln (s)$ is shown in the inset placed on the right, and         
the derivative of  $\ln(s)$ with respect to $\ln(n_{s})$ is shown in the 
inset placed on the left.}
\label{masa3}
\end{figure}

Finally, the same procedure can be applied to understand the oscillation observed 
around the power laws of the size distribution. In fact, figure \ref{masa3} shows 
the tree size distribution function for the $SC_{b}(3,5,5)$ substrate. 
The derivative of $\ln(n_{s})$ with respect to $\ln (s)$ is shown in the inset 
placed on the left-hand side, while the derivative of $\ln(s)$ with respect 
to $\ln(n_{s})$ is shown in the inset placed on the right-hand side.

Based on the obtained results, now we conjecture that in the case of a fractal 
substrate where $k$ tends towards infinity, the logarithmic derivative of the 
size distribution with respect to $\ln(s)$ should be a periodic function around 
a constant value. So, in this case one has

\begin{equation}  
\frac{\partial \ln(n_s)}{\partial \ln(s)} = \tau + \theta (\ln(s)) ,
\label{eq c001} 
\end{equation}

\noindent where $\theta (x)$ is a periodic function.
Then, after integration one gets 

\begin{equation}  
n_s = C s^{\tau} \exp^{\theta^{*} (\ln(s))} ,   
\label{c002} 
\end{equation}

\noindent where $C$ is a constant and $\theta (x) ^{*}$ is a periodic function 
such that $ \partial \theta^{*} (\ln(s)) / \partial \ln(s) = \theta (\ln(s))$.   
Now, by using equation (\ref{c002}) and following the same method 
as in the case of the rms width, it is easy to find that 

\begin{equation}  
\lambda_{n_s} = \tau  \lambda_{s}.
\label{c003} 
\end{equation}

Table II summarizes the values of the different logarithmic periods obtained for the BD model 
on $SC_{x}(3,5,5)$ substrates ($x=a, b, c$). The results were obtained by fitting the data 
shown in the insets of figures 12, 14 and 15, with the aid of equation 
\ref{ajuste.new} for $N=2$.
In the case of the BD model
on $SC_{a}(3,5,5)$ fractals, the data exhibit some fluctuations and it
is no longer
possible to determine a reliable value of $\lambda_{w_s}$. The same
shortcoming is
found when attempting the evaluation of $\lambda_{h_s}$ for the case of the
$SC_{c}(3,5,5)$ substrate. In order to overcome this shortcoming, it
should be necessary
to perform the simulations on fractals with $k \gg 6 $, but this task is
very CPU-time demanding, so it is beyond our computational
capabilities.

\vspace{0.3cm}
\begin{table}[tbp]
\label{TableII}
\caption{List of the logarithmic periods obtained by fitting the 
log-periodic modulations of the power laws describing the internal structure 
of ballistic aggregates on different fractals. The data are also compared
to scaling relationships that are explained in the text.}
\vspace{1.0cm}
{\centering \begin{tabular}{|c|c|c|c|c|c|c|c|c|c|c|c|c|}
\hline 
Fractal  &
\( \lambda_{w_s}    \)&
\( \ln (l)           \)&
\( \lambda_{s}      \)&
\( \lambda_{w_s}/\nu_{\perp}    \)&
\( \lambda_{h_s}    \)&
\( z \ln (l)           \)&
\( \nu_{\parallel} \lambda_{s}      \)&
\( \lambda_{n_s}    \)  &
\( \tau  \lambda_{s} \)    \\
\hline 

\hline 
$SC_{a}(3,5,5)$&
$-$&
$1.0986$&
$3.4(1)$ &
$-$&
$1.80(8)$&
$1.75(6)$&
$1.84(6)$&
$4.8(2)$&
$5.0(2)$      \\

\hline 
$SC_{b}(3,5,5)$&
$1.10(3)$&
$1.0986$&
$3.2(1)$ &
$3.2(1)$ &
$1.70(8)$&
$1.76(6)$&
$1.77(6)$&
$4.7(2)$&
$4.7(2)$      \\

\hline 
$SC_{c}(3,5,5)$&
$1.10(3)$&
$1.0986$&
$3.3(1)$&
$3.2(1)$&
$-$&
$1.77(6)$&
$1.84(6)$&
$4.7(2)$&
$4.9(2)$ \\

\hline 
\hline 
\end{tabular}\par}
\end{table}
\vspace{1.0cm}
  
It is worth mentioning that all determined logarithmic periods can be 
compared to scaling relationships also involving exponents determined 
independently. In fact, the values of $ \lambda_{w_s} $ ($2^{\text nd}$ column) 
are compared to $\ln(l)$ ($3^{\text rd}$ column), as it follows from equation
(\ref{a004}), obtaining an excellent agreement. Also, according to 
equation (\ref{a003}), the period $\lambda_{s}$ ($4^{\text th}$ column) can be 
compared with the quotient given by $\lambda_{w_s}/\nu_{\perp}$    
($5^{\text th}$ column), where the exponent $\nu_{\perp}$ is taken from 
Table I, leading to an excellent agreement.
On the other hand, the period $\lambda_{h_s}$ ($6^{\text th}$ column) can be compared 
with the relationships given by equations (\ref{b004}) ($7^{\text th}$ column)
and (\ref{b003}) ($8^{\text th}$ column), where the exponents $z$ and $\nu_{\parallel}$
are taken from Table I, obtaining agreement within error bars.
Finally, the relationship $\lambda_{h_{ns}} = \tau \lambda_{s}$
given by equation (\ref{c003}), can be verified by comparing columns
$9^{\text th}$ and $10^{\text th}$, also observing agreement within error bars. 

Finally, we would like to comment that for growing aggregates in 
$d-$dimensional homogeneous media, one expects that the average 
volume of the frozen trees of size $s$ ($v_{s}$), will scale 
according to \cite{trees}

\begin{equation}  
v_s \propto h_s w_{s}^{d - 1} \propto  s^{\nu_{\parallel} +(d -1)\nu_{\perp}},   
\label{volume} 
\end{equation} 

\noindent so that one can define $v_s \propto s^{\pi}$, where $\pi$ is an exponent. 
In previous work we showed that for ballistic aggregates in $1 \le d \le 5$
one has that $\pi \simeq 1$, indicating that the trees are compact 
structures \cite{trees}. By using this procedure, we determined 
$\pi = 1.05(2)$, $1.06(2)$, and $1.07(2)$ for the fractals of type 
a, b, and c (see figure \ref{carpetas}), respectively. Since we evaluate
the statistical error only, these exponents could be considered as preliminary
evidence that frozen trees are also compact for ballistic aggregates in 
fractal media.  
 
\section{Conclusions}

We studied the ballistic deposition growth model in deterministic fractal
substrates, embedded in $d = 2$ dimensions, by means of numerical simulations,
analyzing the data according to well established scaling theories.

While the standard Family-Vicsek scaling approach seems to hold at 
least qualitatively, i.e., the interface width exhibits an initial power-law
increase followed by saturation, systematic deviations from the expectation
$W_{sat}(L) \propto L^{\alpha}$ are observed for large $L-$values. In fact, 
the measured values of the interface width are larger than the scaling 
predictions, suggesting that the standard scaling approach has to be modified
in order to properly describe the behavior of the interface upon growth on 
fractal media. Anyway, the obtained (effective) scaling exponents $\alpha$,
$\beta$ and $z$, interpolate between the already known values corresponding to
$d = 1$ and $d = 2$ Euclidean dimensions. Furthermore, the relationship
$z = \alpha/ \beta$ holds, within error bars, and the determined dynamic exponent  
$z$ is in reasonable agreement with subsequent independent measurements, 
performed by studying the internal structure of the aggregates (see, e.g., Table I).   

On the other hand, the study of the properties of the internal structure of the
aggregates, in terms of scaling laws describing the properties of frozen trees,
i.e., branched structures that are inhibited for further growth due to shadowing 
effects and lie below the interface, provides an interesting and rich physical 
behavior. In fact, the standard scaling theory applied to the deposition on
substrates of integer dimension predicts the power-law behavior of relevant
physical observables as a function of the size of the trees, such as the 
rms width and height of the trees, as well as for the tree size distribution.
However, for fractal media with  $d_{f}=\ln(5)/\ln(3)$ and  $d_{f} = \ln(3)/\ln(2)$,
we show that these power laws become modulated by soft log-periodic 
oscillations. These oscillations reflect the interplay between the spatial
discrete scale invariance (DSI) of the underlying fractal substrates and
the growing process. While it is well known that dynamic observables become 
coupled to the {\it spatial} DSI given {\it time} DSI \cite{we2,we,lett},
growing aggregates in fractal media provide a richer scenario. On the one hand,
the rms width of the frozen trees, i.e., an intrinsically spatial observable
measured along the direction parallel to the substrate,
exhibits {\it spatial} DSI with the same fundamental scaling ratio 
($\ln \lambda_{w_{s}} = \ln(l)$ ) as that of the fractal. 
On the other hand, the rms height of the frozen trees, which is 
measured along the growing (time) direction perpendicular to the substrate, 
exhibits {\it time} DSI and its fundamental scaling 
ratio ($\ln \lambda_{h_{s}}$ ) is coupled to 
$\ln \lambda_{w_{s}}$ through the dynamic exponent $z$ (see equation (\ref{b004})),
which is the realization of the more general relationship, also valid for 
any dynamic observable, given by equation (\ref{zeta}). Of course, the 
size distribution of the frozen trees also exhibits DSI, which in a subtle way 
involves both {\it spatial} and {\it time} DSI, since it is an observable 
that depends on the structure of the trees developed along the 
directions parallel and perpendicular to the substrate.      

We hope that this work will contribute to the understanding of growing 
aggregates in fractal media, also stimulating theoretical work aimed 
to describe the interplay between the underlying symmetries of the    
substrates and the behavior of physical observables characteristic
of the processes occurring in those media.
   
{\bf  ACKNOWLEDGEMENTS}. This work was financially supported by 
CONICET, UNLP and  ANPCyT (Argentina).

\end{document}